
\input phyzzx
%
%
\def\abstract{\vskip\frontpageskip
  \centerline{\bf {\titlestyle Abstract}} \vskip\headskip}
\def\ack{\par\penalty-100\medskip \spacecheck\sectionminspace
  \line{\hfil {\titlestyle {\bf Acknowledgements}} \hfil}
  \nobreak\vskip\headskip}
\def\refout{\par \penalty-400 \vskip\chapterskip
  \spacecheck\referenceminspace \immediate\closeout\referencewrite
  \referenceopenfalse
  \line{\hfil {\titlestyle {\bf References}} \hfil}\vskip\headskip
  \input \jobname.refs}

\def\chapter#1{\par \penalty-300 \vskip\chapterskip
  \spacecheck\chapterminspace
  \chapterreset \titlestyle{{\bf \chapterlabel \ \ #1}}
  \nobreak\vskip\headskip \penalty 30000
  \wlog{\string\chapter\ \chapterlabel}}
%
%
\def\NPrefs{\let\therefmark=\NPrefmark \let\therefitem=\NPrefitem}
\def\NPrefmark#1{[#1]}
\def\NPrefitem#1{\refitem{[#1]}}
\NPrefs
%
%
\def\e{\, {\rm e}}
\def\Z{{\bf Z}}
\def\R{{\bf R}}
%
%
%
\REFS\brka{E. Br\'ezin and V.A. Kazakov,
           {\it Phys.\ Lett.\ }{\bf 236B} (1990) 144;
           M. Douglas and S. Shenker,
           {\it Nucl.\ Phys.\ }{\bf B335} (1990) 635;
           D.J. Gross and A.A. Migdal,
           {\it Phys.\ Rev.\ Lett.\ }{\bf 64} (1990) 127;
           {\it Nucl.\ Phys.\ }{\bf B340} (1990) 333.}
\REFSCON\gmil{D.J. Gross and N. Miljkovi\'c,
           {\it Phys.\ Lett.\ }{\bf 238B} (1990) 217;
           E. Br\'ezin, V.A. Kazakov and A. Zamolodchikov,
           {\it Nucl.\ Phys.\ }{\bf B338} (1990) 673;
           P. Ginsparg and J. Zinn-Justin,
           {\it Phys.\ Lett.\ }{\bf 240B} (1990) 333;
           G. Parisi, {\it Phys.\ Lett.\ }{\bf 238B} (1990) 209, 213;
           J. Ambj\o rn, J. Jurkiewicz and A. Krzywicki,
           {\it Phys.\ Lett.\ }{\bf 243B} (1990) 373;
           D.J. Gross and I.R. Klebanov,
           {\it Nucl.\ Phys.\ }{\bf B344} (1990) 475;
           {\it Nucl.\ Phys.\ }{\bf B354} (1991) 459.}
\REFSCON\dika{J. Distler and H. Kawai,
           {\it Nucl.\ Phys.\ }{\bf B321} (1989) 509;
           J. Distler, Z. Hlousek and H. Kawai,
           {\it Int.\ J. of Mod.\ Phys.\ }{\bf A5} (1990) 391; 1093;
           F. David, {\it Mod.\ Phys.\ Lett.\ }{\bf A3} (1989) 1651.}
\REFSCON\seirev{N. Seiberg,
           {\it Prog.\ Theor.\ Phys.\ Suppl.\ }{\bf 102} (1990) 319;
           J. Polchinski, in {\it Strings '90}, eds. R. Arnowitt
           et al., (World Scientific, Singapore, 1991) p.\ 62;
           {\it Nucl.\ Phys.\ }{\bf B357} (1991) 241;
           Y. Kitazawa, Harvard preprint HUTP--91/A034 (1991),
           to appear in {\it Int.\ J. of Mod.\ Phys.\ }{\bf A}.}
\REFSCON\sata{N. Sakai and Y. Tanii,
           {\it Int.\ J. of Mod.\ Phys.\ }{\bf A6} (1991) 2743;
           I.M. Lichtzier and S.D. Odintsov,
           {\it Mod.\ Phys.\ Lett.\ }{\bf A6} (1991) 1953.}
\REFSCON\berkl{M. Bershadsky and I.R. Klebanov,
           {\it Phys.\ Rev.\ Lett.\ }{\bf 65} (1990) 3088;
           {\it Nucl.\ Phys.\ }{\bf B360} (1991) 559.}
\REFSCON\gouli{M. Goulian and M. Li,
           {\it Phys.\ Rev.\ Lett.\ }{\bf 66} (1991) 2051;
           A. Gupta, S. Trivedi and M. Wise,
           {\it Nucl. Phys.\ }{\bf B340} (1990) 475.}
\REFSCON\dfku{P. Di Francesco and D. Kutasov,
          {\it Phys.\ Lett.\ }{\bf 261B} (1991) 385;
           Princeton preprint PUPT--1276 (1991).}
\REFSCON\kita{Y. Kitazawa,
          {\it Phys.\ Lett.\ }{\bf 265B} (1991) 262.}
\REFSCON\sataco{N. Sakai and Y. Tanii,
          {\it Prog.\ Theor.\ Phys.\ }{\bf 86} (1991) 547.}
\REFSCON\dotsenko{V.S. Dotsenko,
          Paris preprint PAR--LPTHE 91--18 (1991).}
\REFSCON\polch{J. Polchinski,
          {\it Nucl. Phys.\ }{\bf B324} (1989) 123;
          {\it Nucl.\ Phys.\ }{\bf B346} (1990) 253;
          S.R. Das, S. Naik and S.R. Wadia,
          {\it Mod.\ Phys.\ Lett.\ }{\bf A4} (1989) 1033;
          S.R. Das and A. Jevicki,
          {\it Mod.\ Phys.\ Lett.\ }{\bf A5} (1990) 1639.}
\REFSCON\grklne{D.J. Gross, I.R. Klebanov and M.J. Newman,
          {\it Nucl.\ Phys.\ }{\bf B350} (1991) 621;
          D.J. Gross and I.R. Klebanov,
          {\it Nucl.\ Phys.\ }{\bf B352} (1991) 671;
          {\it Nucl.\ Phys.\ }{\bf B359} (1991) 3;
          G. Moore, Yale and Rutgers preprint YCTP--P8--91,
          RU--91--12 (1991);
          A.M. Sengupta and S.R. Wadia,
          {\it Int.\ J. of Mod.\ Phys.\ }{\bf A6} (1991) 1961;
          G. Mandal, A.M. Sengupta and S.R. Wadia,
          {\it Mod.\ Phys.\ Lett.\ }{\bf A6} (1991) 1465;
          J. Polchinski, {\it Nucl.\ Phys.\ }{\bf B362} (1991) 125;
          U.H. Danielsson and D.J. Gross,
          {\it Nucl.\ Phys.\ }{\bf B366} (1991) 3.}
\REFSCON\polyakov{A.M. Polyakov,
          {\it Mod.\ Phys.\ Lett.\ }{\bf A6} (1991) 635.}
\REFSCON\satafact{N. Sakai and Y. Tanii,
          Tokyo Inst.\ of Tech.\ and Saitama preprint
          TIT/HEP--173, STUPP--91--120 (1991),
          to appear in {\it Phys.\ Lett.\ }{\bf B};
          TIT/HEP--179,  STUPP--91--122 (1991);
          D. Minic and Z. Yang,
          Texas preprint UTTG--23--91 (1991).}
\REFSCON\lian{B.H. Lian and G.J. Zuckerman,
          {\it Phys.\ Lett.\ }{\bf 266B} (1991) 21;
          S. Mukherji, S. Mukhi and A. Sen,
          {\it Phys.\ Lett.\ }{\bf 266B} (1991) 337;
          P. Bouwknegt, J. McCarthy and K. Pilch, CERN preprint
          CERN--TH--6162--91 (1991).}
\REFSCON\ohta{K. Itoh and N. Ohta,
          Fermilab preprint FERMILAB--PUB--91/228--T (1991);
          K. Aoki and E. D'Hoker, UCLA preprint UCLA--91--TEP--33 (1991);
          E. Abdalla, M.C.B. Abdalla, D. Dalmazi and K. Harada,
          Sao Paulo preprint (1991).}
\REFSCON\kitazawa{Y. Kitazawa,
          Tokyo Inst.\ of Tech.\ preprint TIT/HEP--181 (1991).}
\REFSCON\witten{E. Witten, Princeton preprint IASSNS-HEP-91/51 (1991).}
\REFSCON\avan{J. Avan and A. Jevicki,
          {\it Phys.\ Lett.\ }{\bf 266B} (1991) 35;
          Brown preprints \break
          BROWN--HET--824 and BROWN--HET--839 (1991);
          D. Minic, J. Polchinski and Z. Yang,
          Texas preprint UTTG--16--91 (1991);
          G. Moore and N. Seiberg,
          Rutgers and Yale preprint RU--91--29, YCTP--P19--91 (1991);
          S. Das, A. Dhar, G. Mandal and S. Wadia,
          Princeton preprints IASSNS--HEP--91/52 and 91/72 (1991).}
\REFSCON\klpo{I.R. Klebanov and A.M. Polyakov,
          {\it Mod.\ Phys.\ Lett.\ }{\bf A6} (1991) 3273.}
\REFSCON\kmsk{D. Kutasov, E. Martinec and N. Seiberg, Rutgers and
          Princeton preprint RU--91--49, PUPT--1293 (1991);
          I. R. Klebanov, Princeton preprint PUPT--1302 (1991);
          U. Danielsson, Princeton preprint PUPT--1301 (1991).}
\REFSCON\beku{M. Bershadsky and D. Kutasov, Princeton and Harvard
          preprint PUPT--1283, HUTP--91/A047 (1991);
          Y. Tanii and S. Yamaguchi, Saitama preprint STUPP--91--121
          (1991), to appear in {\it Mod.\ Phys.\ Lett.\ }{\bf A}.}
\REFSCON\rose{M.E. Rose, {\it Elementary Theory of Angular Momentum}
          (John Wiley \& Sons, New York, 1957).}
\REFSCON\gool{P. Goddard and D. Olive, in {\it Vertex Operators in
          Mathematics and Physics}, eds.\ J. Lepowsky et al.,
          (Springer, Heidelberg, 1985) p.\ 51;
          {\it Int.\ J. of Mod.\ Phys.\ }{\bf A1} (1986) 303.}
\refsend
%
%
\Pubnum={TIT/HEP--186 \cr STUPP--92--124}       
\titlepage
\title{\bf Coupling of Tachyons and Discrete States
in $c \! = \! 1$ 2-D Gravity}
\author{Yoichiro Matsumura, Norisuke Sakai}
\address{Department of Physics, Tokyo Institute of Technology \break
         Oh-okayama, Meguro, Tokyo 152, Japan}
\andauthor{Yoshiaki Tanii}
\address{Physics Department, Saitama University \break
         Urawa, Saitama 338, Japan}
\abstract
All the three point couplings involving tachyons and/or discrete states
are obtained in $c=1$ two-dimensional (2-D) quantum gravity by
means of the operator product expansion (OPE).
Cocycle factors are found to be necessary in order to maintain the
analytic structure of the OPE, and are constructed explicitly both
for discrete states and for tachyons.
The effective action involving tachyons and discrete states is worked
out to summarize all of these three point couplings.
\endpage
%
%
\chapter{Introduction}
Recent advances in the matrix model \NPrefmark{\brka, \gmil}
for the nonperturbative treatment of two-dimensional
(2-D) quantum gravity has prompted much progress with the
continuum approach by means of the Liouville theory.
In spite of the nonlinear dynamics of the Liouville theory,
a method based on conformal field theory has now been sufficiently
developed to understand the results of the matrix model
and to offer in some cases a more powerful method for computing
various quantities \NPrefmark{\dika, \seirev}.
In particular, one can now calculate not only partition
functions \NPrefmark{\sata, \berkl} but also correlation
functions, using the procedure of analytic
continuation  \NPrefmark{\gouli - \dotsenko}.
\par
So far only conformal field theories with central charge $ c \le 1$
have been successfully coupled to quantum gravity.
The $c = 1$ case is the richest and the most interesting.
It has been observed that $c=1$ quantum gravity can be regarded
effectively as a critical string theory in two
dimensions, since the Liouville field zero
mode provides an additional ``time-like'' dimension besides the
obvious single spatial dimension given by the zero mode of the $c=1$
matter \NPrefmark{\polch}.
The center of mass motion of the string provides a physical scalar
particle.
In the usual bosonic string theory at the critical dimension $(D=26)$,
the scalar particle has negative squared mass and is called tachyon.
In the present case of a noncritical string, the scalar particle
becomes massless but is still called ``tachyon'' following the usual
terminology of critical string theory.
Since there are no transverse directions, the continuous (field)
degrees of freedom are exhausted by the tachyon field.
However, it has been noted that there exist other
discrete degrees of freedom in the $ c=1$ matter coupled to the 2-D
quantum gravity \NPrefmark{\grklne-\kitazawa}.
It has been pointed out that the symmetry group relevant to the
dynamics of these discrete states in the $c=1$ quantum gravity is
that of the area preserving diffeomorphisms whose generators fall into
representations of SU(2) \NPrefmark{\witten, \avan}.
By using the SU(2) symmetry, Klebanov and Polyakov
have recently worked out the three point interactions of the
discrete states and have proposed an effective action for these
discrete states in the case of integer spins \NPrefmark{\klpo}.
\par
The purpose of this paper is to obtain all possible three point
couplings of tachyons and discrete states in $c=1$ quantum gravity,
and to write down an effective action representing all of these
couplings.
Using the operator product expansion (OPE),
we have found all possible three point couplings involving
tachyons and/or discrete states: couplings among three tachyons,
and couplings among two tachyons and a single discrete state
in addition to those among three discrete states which were already
worked out \NPrefmark{\klpo}. We have also found that cocycle factors
are needed to maintain analyticity of the OPE and have constructed
them explicitly. An effective action is worked out for the three point
coupling of discrete states even in the presence of half odd
integer spins. We have also found an effective action representing
the three point coupling involving tachyons.
\par
In the next section, tachyon and discrete state vertex operators are
constructed as physical states of the $c=1$ quantum gravity.
In sect.\ 3, the cocycle factors are shown to be necessary to
maintain proper analytic behaviour of the OPE, and they are
explicitly constructed.
The OPEs involving tachyons are worked out in sect.\ 4.
In sect.\ 5, an effective action is constructed for the discrete states
when there are half odd integer spins beside integer spins.
An effective action for couplings involving tachyons is obtained
in sect.\ 6.
\par
While we are writing this paper, we have received papers
\NPrefmark{\kmsk}\ where the operator algebra involving the discrete
states and the tachyon operators is discussed.
Part of our results has some overlap with theirs,
although our method is different from theirs.
\par
%
%
\chapter{Tachyons and Discrete States}
We consider the $c=1$ conformal matter realized by a single scalar
field (string variable) $X$ coupled to two-dimensional quantum
gravity. In the conformal gauge
$g_{\alpha\beta} = \e^\phi \hat g_{\alpha\beta}$ with
a reference metric $\hat g_{\alpha\beta}$, the Liouville field $\phi$
represents all the local degrees of freedom of the metric.
The dynamics of $X$ and $\phi$ can be
described by the following action  on a surface with boundary
\NPrefmark{\dika, \seirev, \beku}
$$
\eqalign{
S[\hat g, X, \phi]
= \ & {1 \over 4\pi\alpha'} \int d^2 z \sqrt{\hat g}
\Bigl( \hat g^{\alpha\beta} \partial_\alpha X \partial_\beta X
+ \hat g^{\alpha\beta} \partial_\alpha \phi \partial_\beta \phi
- 2\sqrt{\alpha'} \hat R \phi \cr
& + 4\alpha' \mu \e^{-2 \phi / \sqrt{\alpha'}} \Bigr)
+ {1 \over \pi\sqrt{\alpha'}} \int d \hat s
\left( - \hat k \phi + \sqrt{\alpha'} \lambda
\e^{-\phi / \sqrt{\alpha'}} \right),
}\eqn\action
$$
where $\alpha'$ is the Regge slope parameter, $\hat R$ the scalar
curvature, $\hat k$ the geodesic curvature along the boundary
and $d \hat s$ the line element of the boundary with respect to
the reference metric $\hat g_{\alpha\beta}$.
We have rescaled the Liouville field $\phi$.
The action \action\ can be regarded as describing a critical string
theory in a two-dimensional target space-time  with ``time''
$\phi$ and a spatial coordinate $X$ \NPrefmark{\polch}.
Since the large negative values
of the Liouville field $\phi$ are suppressed by the
cosmological term even for a small cosmological constant $\mu$,
the zero mode of $\phi$ is effectively limited to a finite
length proportional to $\ln\mu$.
Amplitudes generally have contributions proportional to $\ln\mu$
which are called bulk or resonant amplitudes \NPrefmark{\polyakov}.
In the following, we will consider only the bulk correlation
functions, for which the ``energy'' and
the momentum conjugate to $\phi$ and $X$ respectively are conserved.
To compute such correlation functions we can put  $\mu = \lambda = 0$.
Consequently we shall use the action without the cosmological terms
in the following, and compute the correlation functions
by means of the conformal field theory technique with free field
realizations \NPrefmark{\gouli-\dotsenko, \satafact}.
\par
There are two types of physical operators.
The open string vertex operators are given by line integrals of
primary fields with boundary conformal weight one along the boundary,
while the closed string vertex operators are given by surface
integrals of primary fields with conformal weight $(1, 1)$.
It is convenient to set $\alpha' = 4 \; (1)$ when we discuss
the closed (open) string vertex operators. With this convention
the integrands of the closed string vertex operators
can be constructed by combining the holomorphic operator and the
anti-holomorphic operator, both of which have the same form as those of
the open string vertex operators.
\par
Let us first consider the open string vertex operators.
The simplest field for such operators is the
gravitationally dressed tachyon vertex operator with momentum $p$
$$
\Psi^{(\pm )}_p (z) = \e^{ipX(z)} \e^{(\pm p-1)\phi(z)}.
\eqn\tachyon
$$
The requirement that conformal weight be unity gives two solutions
$\pm p-1$ for the Liouville energy.
The upper (lower) sign is called positive (negative)
chirality \NPrefmark{\polyakov}.
One should note that the
momentum $p$ can take arbitrary real values if the boson $X$
is noncompact.
For higher levels there are non-trivial primary fields
only when the momentum is an integer or a half odd integer.
They are primary fields for
the ``discrete states'' \NPrefmark{\grklne, \polyakov}.
They form SU(2) multiplets and can be constructed
as \NPrefmark{\witten, \klpo}
$$
\Psi^{(\pm )}_{J,m} (z) = \sqrt{(J+m)! \over (2J)! (J-m)!}
\oint {du_{J-m} \over 2\pi i} H_-(u_{J-m}) \cdots
\oint {du_1 \over 2\pi i} H_-(u_1) \Psi^{(\pm )}_{J} (z),
\eqn\discrete
$$
where $J = {1 \over 2}, 1, \cdots \; $; $m= -J, -J+1, \cdots, J$
and $\Psi^{(\pm )}_{J} (z)$ is the tachyon operator \tachyon\ with
the momentum $p=J$. The integrals are along closed contours
surrounding a point $z$ with $|u_i| > |u_j|$ for $i > j$.
The field $H_-(z)$ corresponds to the lowering operator of the
SU(2) quantum numbers and is one of the SU(2) currents
$$
H_\pm (z) = \e^{\pm i X(z)} = \pm \Psi^{(+)}_{1, \pm 1} (z), \quad
H_3 (z) = {1 \over 2} i \partial X(z)
= -{1 \over \sqrt 2} \Psi^{(+)}_{1, 0} (z).
\eqn\sugenerator
$$
The quantum numbers $J$ and $m$ correspond to the ``spin'' and the
magnetic quantum number in SU(2).
Actually, the fields $\Psi^{(\pm )}_{J, m}$ with $m= \pm J$ are not
higher level operators but tachyon operators  \tachyon\ at integer
or half odd integer momenta $\pm J$.
$$
\Psi^{(\pm )}_{J, J}(z) = \Psi^{(\pm )}_J (z), \quad
\Psi^{(\pm )}_{J, -J}(z) = (-1)^{J(2J-1)} \Psi^{(\mp)}_{-J} (z).
\eqn\distach
$$
The sign factor in the second equation is a consequence of the
definition \discrete\ which respects the usual relative sign
convention of states with different $m$ \NPrefmark{\rose}.
It should be noted that the type of discrete state $(+)$ or
$(-)$ is opposite to the chirality of the corresponding tachyon
for the case of lowest magnetic quantum number $m=-J$.
\par
Strictly speaking, the expression \discrete\ is correct only for
integer $J$ as we will see in the next section.
We will explain how to modify eq.\ \discrete\ to give a correct
expression for half odd integer $J$ when we discuss cocycle operators
in the next section.
\par
%
\chapter{Discrete State OPE and Cocycle}
In ref.\ \NPrefmark{\klpo} the OPEs of the fields for discrete
states \discrete\ were obtained using the SU(2) symmetry.
Here we make a remark on the analytic property of the OPEs.
The OPE of two vertex operators in eq.\ \discrete\ at $z$ and $w$
gives a result different in sign depending on the ordering of
the two vertex operators.
Hence the OPE is not analytic in the complex coordinates $z$ at
$|z| = |w|$, even if we use the
radial ordering of the two vertex operators
as usual in conformal field theory.
To be more precise, let us consider the OPE of
$\Psi^{(+)}_{J_1, m_1} (z)$ and $\Psi^{(+)}_{J_2, m_2} (w)$.
The conservation of momentum and the Liouville energy follows from the
zero mode dependence of $X$ and $\phi$.
Hence one immediately finds that the OPE of the two operators gives
the operator $\Psi^{(+)}_{J_1+J_2-1, m_1+m_2}(w)$ as the only possible
operator with conformal weight one.
When $|z| > |w|$, the product of the two operators gives
${1 \over z-w}F_{J_1, m_1 \, J_2, m_2}
\Psi^{(+)}_{J_1+J_2-1, m_1+m_2}(w)$
with a certain coefficient $F_{J_1, m_1 \, J_2, m_2}$.
On the other hand, when $|w| > |z|$, it gives
${1 \over z-w}F'_{J_1, m_1 \, J_2, m_2}
\Psi^{(+)}_{J_1+J_2-1, m_1+m_2}(w)$
with a different coefficient $F'_{J_1, m_1 \, J_2, m_2}$.
By using explicit representations \discrete, it can be shown that
they differ by a sign
$$
F_{J_1, m_1 \, J_2, m_2}
= (-1)^{2J_1(J_2-m_2-1)+2J_2(J_1-m_1-1)} F'_{J_1, m_1 \, J_2, m_2}.
\eqn\sign
$$
Therefore, the coefficient function of the OPE is not analytic
at $|z| = |w|$.
It is desirable to obtain OPEs with analytic coefficients since the
techniques of conformal field theories make full use of the
analyticity. The analytic OPE can be achieved by multiplying
a correction factor to the operators \discrete\ as in the vertex
operator construction of the affine Kac-Moody
algebra \NPrefmark{\gool}.
\par
The correction factor can be constructed as follows.
Let us consider a two-dimensional lattice
$$
\Lambda = \left\{ \alpha = \sqrt 2 (m, J-1) \Bigm\vert
m, J \in {1 \over 2} \Z, \; J-m \in \Z \right\}.
\eqn\lattice
$$
with a Lorentzian inner product
$$
\alpha_1 \cdot \alpha_2 = 2 m_1 m_2 - 2 (J_1-1) (J_2-1), \quad
\alpha_1, \;\, \alpha_2 \in \Lambda.
\eqn\innerprod
$$
The lattice $\Lambda$ is an even lattice, i.e.
$\forall \alpha \in \Lambda, \; \alpha \cdot \alpha \in 2\Z$ and
therefore it is an integral lattice: $\forall \alpha, \; \beta \in
\Lambda, \; \alpha \cdot \beta \in \Z$.
Furthermore, the lattice $\Lambda$ is self-dual
$\Lambda = \Lambda^\ast$,
where $\Lambda^\ast = \{ x \, | \, \forall \alpha \in \Lambda, \;
\alpha \cdot x \in \Z \}$ is the dual lattice of $\Lambda$.
Using these notations the sign factor in eq.\ \sign\ can be written
as $(-1)^{\alpha_1 \cdot \alpha_2}$. The effect of the correction
factor must cancel this sign factor.
\par
To construct the correction factor we need a cocycle
$\varepsilon (\alpha, \beta) = \pm 1 \; (\alpha, \beta \in \Lambda)$
which satisfies
$$
\eqalign{
\varepsilon (\alpha, \beta)
& = (-1)^{\alpha \cdot \beta} \varepsilon (\beta, \alpha), \cr
\varepsilon (\alpha, \beta) \varepsilon (\alpha + \beta, \gamma)
& = \varepsilon (\alpha, \beta + \gamma) \varepsilon (\beta, \gamma).
}\eqn\cocycle
$$
The cocycle $\varepsilon (\alpha, \beta)$ for $\Lambda$ can be
obtained as follows. Consider a three-dimensional Lorentzian cubic
lattice with the metric $(1, -1, 1)$ for diagonal components
$$
\Lambda_3
= \left\{ \tilde \alpha = \sum_{i=1}^3 n^i \tilde {\bf e}_i
\Bigm\vert n^i \in \Z \right\}
\eqn\cubiclattice
$$
with basis vectors
$$
\tilde {\bf e}_1
= \left( {1 \over \sqrt 2}, -{1 \over \sqrt 2}, 1 \right), \quad
\tilde {\bf e}_2
= \left( -{1 \over \sqrt 2}, -{1 \over \sqrt 2}, 1 \right), \quad
\tilde {\bf e}_3 = \left( 0, -\sqrt 2, 1 \right).
\eqn\basis
$$
The inner product is given by
$$
\tilde {\bf e}_i \cdot \tilde {\bf e}_j = \left\{
\eqalign{
+1 & \quad (i=j=1, 2) \cr
0 & \quad (i \not= j) \cr
-1 & \quad (i=j=3). } \right.
\eqn\basisprod
$$
The two-dimensional lattice $\Lambda$ can be identified with
a sublattice of $\Lambda_3$
$$
\Lambda \sim \left\{ \tilde \alpha = \sum_{i=1}^3 n^i \tilde {\bf e}_i
\Bigm\vert n^i \in \Z, \; n_1 + n_2 + n_3 = 0 \right\}.
\eqn\sublattice
$$
For the cubic lattice $\Lambda_3$ it is easy to obtain the cocycle
factor by using the Clifford algebra associated with the
lattice \NPrefmark{\gool}.
The cocycle factor of $\Lambda$ can be obtained from that
of $\Lambda_3$ by restricting it to the sublattice \sublattice.
Thus we find the following choice of the cocycle factor for $\Lambda$
$$
\varepsilon (\alpha_1, \alpha_2) = (-1)^{2s_1 J_1(s_2 J_2-m_2-1)}
\eqn\cocycleresult
$$
for the two-vector corresponding to the type $s_i(= \pm)$ discrete
states
$$
\alpha_i = \sqrt 2 \, (m_i, s_iJ_i-1), \quad i=1,2.
\eqn\twovector
$$
It is easy to see that eq.\ \cocycleresult\ indeed satisfies the
cocycle conditions \cocycle.
\par
With this cocycle factor we can construct the correction factor
which is called the cocycle operator \NPrefmark{\gool}
$$
c_\alpha = \sum_{\beta \in \Lambda} \varepsilon (\alpha, \beta)
\ket{\beta} \bra{\beta},
\eqn\calpha
$$
where $\ket{\beta}$ is an eigenstate of the energy and the momentum
with an eigenvalue $\beta / \sqrt 2$ as in eq. \lattice.
Then the corrected operators
$$
\Psi'^{\, (s)}_{J, m} (z) = \Psi^{(s)}_{J, m} (z) \, c_\alpha,
\quad \alpha = \sqrt 2 \, (m, sJ-1)
\eqn\cdiscrete
$$
satisfy the OPEs which are analytic in the complex $z$ plane.
The fields in eqs.\ \discrete\ and \sugenerator\ should also be
replaced by corrected operators with cocylce operators included.
Then the expression \discrete\ is correct for half odd integer $J$
as well as for integer $J$.
We find that after an appropriate rescaling the corrected
operators \cdiscrete\ satisfy the same OPEs as those given in
ref.\ \NPrefmark{\klpo}. The non-trivial OPEs are given by
$$
\eqalign{
\tilde\Psi'^{\,(+)}_{J_1, m_1}(z) \, \tilde\Psi'^{\,(+)}_{J_2, m_2}(w)
& \sim \, {1 \over z-w} \, (J_2 m_1- J_1 m_2) \,
\tilde\Psi'^{\,(+)}_{J_1+J_2-1, m_1+m_2}(w), \cr
\tilde\Psi'^{\,(+)}_{J_1, m_1}(z) \,
\tilde\Psi'^{\,(-)}_{J_1+J_3-1, -m_1+m_3}(w)
& \sim \, {1 \over z-w} \, (- J_1 m_3 - J_3 m_1) \,
\tilde\Psi'^{\,(-)}_{J_3, m_3}(w).
}\eqn\discreteope
$$
Other OPEs have no singular term.
We have used rescaled fields
$$
\eqalign{
\tilde\Psi'^{\,(+)}_{J,m}(z) &
= \tilde N(J, m) \, \Psi'^{\,(+)}_{J,m}(z), \cr
\tilde\Psi'^{\,(-)}_{J,m}(z) &
= (-1)^{J(2J-1)+J-m} \left[ \tilde N(J, m) \right]^{-1}
\Psi'^{\,(-)}_{J,m}(z),
}\eqn\rescale
$$
where
$$
\tilde N(J, m) = (2J-1)! \sqrt{J \over 2} N(J, m), \quad
N(J, m) = \left[ {(J+m)!(J-m)! \over (2J-1)!} \right]^{1 \over 2}.
\eqn\njm
$$
\par
%
%
\chapter{Tachyon OPE}
We shall now generalize these results for the OPE to include the
tachyon operator \tachyon.
Since the discrete state OPE \discreteope\ contains tachyon states
with integer or half odd integer momenta as in eq.\ \distach,
it is clear that the tachyon OPE should also have cocycle factors
in order to maintain analyticity.
Although the construction of cocycle factors for a tachyon vertex
operator with arbitrary momentum is difficult or perhaps
ill-defined, we can construct cocycle factors restricted for our
purposes.
As we see below, we need to discuss the tachyon OPE in the case of
two momenta adding up to integer or half odd integer values.
Let us define the integer or half odd integer part $J$ of
momentum $p$ as
$$
J = \left\{\eqalign{
n              & \qquad {\rm for}\;\,  n < p < n+{1 \over 4} \cr
n+{1 \over 2}  & \qquad {\rm for}\;\,
n+{1 \over 4} < p < n+{3 \over 4} \cr
n+1            & \qquad {\rm for}\;\, n+{3 \over 4} < p < n+1 } \right.
\eqn\jhalfint
$$
for $n \in \Z$. From this definition we immediately find
$$
J_1+J_2=p_1+p_2 \qquad {\rm if} \qquad
p_1+p_2 \in \Z \quad {\rm or} \quad \Z+{1 \over 2}.
\eqn\sumhalfint
$$
We shall define the integer or half odd integer part $J$ for
$p=n+{1 \over 4}$ or $p=n+{3 \over 4}$ as the limit of the above
cases. Therefore two momenta adding up to integer or half odd integer
actually correspond to the limit of one of the momenta approaching from
above and the other from below. We can now associate a two-vector
$\alpha$ on the lattice for the tachyon with momentum $p$ whose
integer or half odd integer part is $J$ and whose chirality is $s$
$$
\alpha = \sqrt 2 \, (J, \; sJ-1).
\eqn\tachtwovect
$$
These two-vectors belong to the lattice \lattice, and allow us to
use the same cocycle factor $\varepsilon (\alpha, \beta)$
in eq.\ \cocycleresult.
Since the cocycle factor satisfies the same algebra as before,
it satisfies the cocycle condition.
We can now define the cocycle operator for the tachyon with momentum
$p$ using the corresponding two-vector $\alpha$ as
$$
c_p = \sum_{\beta \in \Lambda} \varepsilon (\alpha, \beta)
\ket{\beta+\alpha-\hat p} \bra{\beta}, \qquad
\hat p= \sqrt 2 \, (p, s p-1),
\eqn\cp
$$
where the two-vector $\hat p$ is to cancel the effect of multiplying
the zero mode part of the vertex operator $\e^{ipX_0+(sp-1)\phi_0}$.
More precisely speaking, we can add any fixed momentum $\bar p$ to
all of the momentum eigenstates appearing in the cocycle operator.
Precisely just as for the ordinary lattice for various Lie
algebras \NPrefmark{\gool}, this momentum $\bar p$ just specifies
the conjugacy class of momenta: the conjugacy class in the present
case is defined by the relation that two momenta add up to give half
integer values. We can choose $0 < \bar p< {1 \over 4}$.
In fact, the cocycle operators are only effective for two momenta from
the same conjugacy class.
With this cocycle operator, we can construct the corrected operator
$$
\Psi'^{\, (s)}_{p} (z) = \Psi^{(s)}_{p} (z) c_\alpha,
\eqn\ctachyon
$$
which can be shown to satisfy the OPEs with analytic coefficient
functions.
\par
{}From the conservation of energy and momentum
we find that only four non-trivial OPEs are possible:
$$
\eqalign{
\Psi'^{(+)}_{p_1}(z) \, \Psi'^{(+)}_{p_2}(w)
& \sim \, {1 \over z-w} \, F^{(+)}_{p_1 p_2} \,
\tilde\Psi'^{(-)}_{J_3, 1-J_3}(w) \quad (J_3 = -p_1-p_2+1), \cr
\Psi'^{(-)}_{p_1}(z) \, \Psi'^{(-)}_{p_2}(w)
& \sim \, {1 \over z-w} \, F^{(-)}_{p_1 p_2} \,
\tilde\Psi'^{(-)}_{J_3, J_3-1}(w) \quad (J_3 = p_1+p_2+1), \cr
\tilde\Psi'^{(+)}_{J_1, J_1-1}(z) \, \Psi'^{(+)}_{p_2}(w)
& \sim \, {1 \over z-w} \, G^{(+)}_{J_1 p_2} \,
\Psi'^{(+)}_{p_3}(w) \quad (p_3 = J_1-1+p_2), \cr
\tilde\Psi'^{(+)}_{J_1, 1-J_1}(z) \, \Psi'^{(-)}_{p_2}(w)
& \sim \, {1 \over z-w} \, G^{(-)}_{J_1 p_2} \,
\Psi'^{(-)}_{p_3}(w) \quad (p_3 = 1-J_1+p_2).
}\eqn\tachyonope
$$
Since the discrete states with $m=\pm J$ are actually tachyons, the
above OPEs contain the cases involving three tachyons and without
any discrete state. We find that the above OPEs exhaust all
possible three point couplings involving tachyons.
We have to compute the coefficients $F$'s and $G$'s.
The coefficient in the third OPE in eq.\ \tachyonope\ can easily be
obtained using the representation \discrete\ for
$\Psi'^{(+)}_{J_1, J_1-1}$ and directly evaluating the OPE.
The coefficient in the last OPE in eq.\ \tachyonope\ can be evaluated
similarly using the representation
$$
\Psi'^{(+)}_{J, 1-J} (z) = (-1)^{J(2J-1)} {1 \over \sqrt{2J}}
\oint {du \over 2\pi i} H_+ (u) \Psi'^{(-)}_{-J} (z),
\eqn\jj
$$
which can be derived from eq.\ \discrete. In this way we find
$$
\eqalign{
G^{(+)}_{J_1 p_2}
& =-{\Gamma(1+2p_3) \over 2 \, \Gamma(2p_2)}
= (-1)^{2J_1}{\Gamma(1-2p_2) \over 2 \, \Gamma(-2p_3)}
= -{\tilde N(p_3, p_3) \over \tilde N(p_2, p_2)} \, p_2, \cr
G^{(-)}_{J_1 p_2}
& =(-1)^{J_2(2J_2-1)+J_3(2J_3-1)+1}
{\Gamma(1+2p_2) \over 2 \, \Gamma(2p_3)} \cr
& =(-1)^{J_2(2J_2-1)+J_3(2J_3-1)+2J_1}
{\Gamma(1-2p_3) \over 2 \, \Gamma(-2p_2)} \cr
&= (-1)^{J_2(2J_2-1)+J_3(2J_3-1)+1} \,
{\tilde N(p_2, p_2) \over \tilde N(p_3, p_3)} \, p_3,
}\eqn\opegco
$$
where $\tilde N(p, p) = {1 \over 2} \Gamma (1+2p)$.
One should note that the sign factors due to the cocycle
$(-1)^{2J_1}$ and $(-1)^{2J_1(2J_2-1)}$ for the first and the second
equations respectively have been included in the above formulas.
\par
To obtain the coefficient of the first OPE in eq.\ \tachyonope, we
apply the operator $\oint {du \over 2\pi i} H_- (u)$ to both hand sides
of the equation, where the integration contour surrounds both of
$z$ and $w$. The left hand side of the OPE becomes
$$
\oint {du \over 2\pi i} H_- (u)
\Psi'^{(+)}_{p_1}(z) \Psi'^{(+)}_{p_2}(w)
\sim {1 \over z-w} \oint {dx \over 2\pi i} x^{-2p_1} (1+x)^{-2p_2}
\Psi'^{(-)}_{J_3, -J_3}(w),
\eqn\lhs
$$
where we have changed the integration variable to $x=(u-z)/z$
and the $x$-integration surrounds $0$ and $-1$.
The right hand side of the OPE becomes proportional to
$$
\oint {du \over 2\pi i} H_- (u) \Psi'^{(-)}_{J_3, 1-J_3}(w)
= \sqrt{2J_3} \, \Psi'^{(-)}_{J_3, -J_3}(w).
\eqn\rhs
$$
Therefore, by evaluating the integral in eq.\ \lhs\ we obtain
the coefficient $F^{(+)}$.
The coefficient of the second OPE in
eq.\ \tachyonope\ can be obtained similarly by applying
$\oint {du \over 2\pi i} H_+ (u)$.
We find
$$
\eqalign{
F^{(+)}_{p_1 p_2}
& = (-1)^{2J_1}{\Gamma(1-2p_1) \over 2\, \Gamma(2p_2)}
  = (-1)^{2J_2-1}{\Gamma(1-2p_2) \over 2 \, \Gamma(2p_1)} \cr
& = (-1)^{2J_1}\left[ \tilde N(p_1, p_1)
    \tilde N(p_2, p_2) \right]^{-1}
    {\pi p_1 p_2 \over 2 \sin (2 \pi p_1)}, \cr
F^{(-)}_{p_1 p_2}
& = (-1)^{J_1(2J_1-1)+J_2(2J_2-1)-2J_1}
    {\Gamma(1+2p_2) \over 2 \, \Gamma(-2p_1)} \cr
& = (-1)^{J_1(2J_1-1)+J_2(2J_2-1)+2J_2-1}
    {\Gamma(1+2p_1) \over 2 \, \Gamma(-2p_2)} \cr
& = (-1)^{J_1(2J_1-1)+J_2(2J_2-1)+2J_1-1} \tilde N(p_1, p_1)
    \tilde N(p_2, p_2)  {2\sin (2\pi p_1) \over \pi}.
}\eqn\opefco
$$
We have included the sign factor due to the cocycle
$(-1)^{2J_1}$ and $(-1)^{2J_1(2J_2-1)}$ for the first and the second
equations respectively.
\par
%
%
\chapter{Effective Action for Discrete States}
The coefficients of the OPE determine the three-point correlation
functions of the physical operators. From SL(2, \R) symmetry
the three-point function of the fields takes the form
$$
\VEV{\tilde\Psi'^{(+)}_{J_1, m_1}(z_1) \,
\tilde\Psi'^{(+)}_{J_2, m_2}(z_2) \,
\tilde\Psi'^{(-)}_{J_3, m_3}(z_3)}
= {c(J_1, m_1; J_2, m_2; J_3, m_3) \over
(z_1-z_2)(z_2-z_3)(z_3-z_1)}.
\eqn\threepoint
$$
Therefore the three-point function of the integrated physical
operators is given by the coefficient
$c(J_1, m_1; J_2, m_2; J_3, m_3)$.
Using the first OPE of eq.\ \discreteope\ in eq.\ \threepoint,
we find that the coefficient is given by a product
of the OPE coefficient and the two-point correlation function.
Let us obtain the two-point function.
{}From the SU(2) and SL(2, \R) symmetries it takes the form
$$
\eqalign{
\VEV{\Psi'^{(+)}_{J_1, m_1}(z) \, \Psi'^{(-)}_{J_2, m_2}(w)}
& = \VEV{J_1 \, J_2 \, m_1 \, m_2 \vert 0 \, 0} \, c(J_1, J_2)
\, {1 \over (z-w)^2} \cr
& = {(-1)^{J_2-m_1} \over \sqrt{2J_2+1}} \, c(J_1, J_2)
\, \delta_{J_1, J_2} \, \delta_{m_1, -m_2} \, {1 \over (z-w)^2}.
}\eqn\twopoint
$$
The coefficient $c(J_1, J_2)$ can be determined by explicitly
evaluating the two-point function for $J_1 = m_1 = J_2 = -m_2$.
Thus we find
$$
\VEV{\Psi'^{(+)}_{J_1, m_1}(z) \, \Psi'^{(-)}_{J_2, m_2}(w)}
= \delta_{J_1, J_2} \, \delta_{m_1, -m_2} \,
{(-1)^{J_2(2J_2-1)+J_2+m_2} \over (z-w)^2}.
\eqn\twopointresult
$$
Using eq.\ \twopointresult\ we find the coefficient of the
three-point function is given by
$$
c(J_1, m_1; J_2, m_2; J_3, m_3) = - (J_2 m_1 - J_1 m_2) \,
\delta_{J_1+J_2-1, J_3} \, \delta_{m_1+m_2+m_3, 0}.
\eqn\threepointresult
$$
\par
The result of the correlation functions can be summarized by the
effective action, which reproduces the three-point function
$c(J_1, m_1; J_2, m_2; J_3, m_3)$ of the integrated physical
operators.
Introducing a variable $g^{(s)}_{J,m}$ $(s=\pm )$
for each discrete state, the cubic terms of the effective action
determined by the OPEs \discreteope\ are \NPrefmark{\klpo}
$$
S_3 = {g_o \over 2} \sum_{J_1, m_1, J_2, m_2, A, B, C}
(J_2 m_1-J_1 m_2) f^{ABC} g^{(-) A}_{J_1+J_2-1, -m_1-m_2}
g^{(+) B}_{J_1, m_1} g^{(+) C}_{J_2, m_2}
 \int d\phi,
\eqn\cubic
$$
where we have introduced the Chan-Paton index $A$ in the adjoint
representation of some Lie algebra and open string coupling constant
$g_o$.
\par
In ref.\ \NPrefmark{\klpo} it was shown that the terms in the
cubic interaction \cubic\ which depend only on the integer modes
$g^{(s) A}_{J, m} \;\; (J, \, m \in \Z)$ can be written in a
compact form by introducing a scalar field on $\R \times {\rm S}^2$
$$
\Phi_0 (\phi, \theta, \varphi)
= \sum_{s, A, J, m} T^A g_{J, m}^{(s) A} M^s(J, m)
D^J_{m \, 0} (\varphi, \theta, 0) \e^{(s J-1)\phi}.
\eqn\integermode
$$
Here, $T^A$ are the representation matrices of the Lie algebra and
$D^J_{m \, 0}$ are components of the SU(2) rotation
matrix \NPrefmark{\rose}
$$
\eqalign{
& D^J_{m \, m'} (\varphi, \theta, \psi)
= \bigl\langle J m \bigr\vert \e^{-i\varphi J_z} \e^{-i\theta J_y}
\e^{-i\psi J_z} \, \bigl\vert \, J m' \bigr\rangle, \cr
& \qquad\qquad 0 \leq \varphi, \psi < 2\pi,
\quad 0 \leq \theta \leq \pi.
}\eqn\rmatrix
$$
The coefficients $M^s (J, m)$ are chosen to be
$$
M^+ (J, m) = {N(J, m) N(J, 0) \over J}, \quad
M^- (J, m) = {(-1)^m \over 4\pi} {J(2J+1) \over N(J, m) N(J, 0)}.
\eqn\integermm
$$
In terms of the field $\Phi_0$ the effective action can be written as
$$
S_3^{(1)} = {1 \over 3}i g_o \int d\phi \e^{2\phi}
\int\nolimits_{S^2} d\theta d\varphi \,
\epsilon^{ij} \, {\rm Tr} \left( \Phi_0
{\partial \Phi_0 \over \partial x^i}
{\partial \Phi_0 \over \partial x^j} \right),
\eqn\integeraction
$$
where $x^i = (\theta, \varphi)$.
\par
We shall generalize this construction to the terms depending
on half odd integer modes as well as integer modes.
We introduce two spinor fields
$\Phi_{1 \over 2}$ and $\Phi_{-{1 \over 2}}$ on
$\R \times {\rm S}^2$ for half odd integer modes
$g_{J, m}^{(s) A} \;\; (J, \, m \in \Z + {1 \over 2})$
$$
\Phi_\mu (\phi, \theta, \varphi)
= \sum_{s, A, J, m} T^A g_{J, m}^{(s) A} M^s_\mu(J, m)
D^J_{m \, \mu} (\varphi, \theta, 0) \e^{(s J-1)\phi}
\quad \left( \mu = \pm {1 \over 2} \right),
\eqn\halfintegermode
$$
where
$$
M^+_\mu (J, m) = {N(J, m) N(J, {1 \over 2}) \over J+{1 \over 2}}, \quad
M^-_\mu (J, m) = {(-1)^{m+\mu} \over 4\pi} {2J(J+1)
\over N(J, m) N(J, {1 \over 2})}.
\eqn\halfintegermm
$$
Note that $\Phi_{1 \over 2}$ and $\Phi_{-{1 \over 2}}$ have the same
coefficients $g_{J, m}^{(s) A}$ and therefore are not independent.
In order to write down the effective action in terms of these fields
we need covariant derivatives on ${\rm S}^2$ acting on spinor fields
$\Phi_{\mu}$. They are given by
$$
\nabla_\pm = \mp \partial_\theta -
{1 \over \sin\theta} (i \partial_\varphi - \mu \cos\theta)
\eqn\covd
$$
when acting on $\Phi_\mu$. They act on the rotation matrix as
raising and lowering operators:
$$
\nabla_\pm D^J_{m \, \mu} (\varphi, \theta, 0)
= \sqrt{(J \mp \mu)(J \pm \mu + 1)} \,
D^J_{m \, \mu \pm 1} (\varphi, \theta, 0).
\eqn\dd
$$
We also need an integration formula for the rotation
matrix \NPrefmark{\rose}
$$
\eqalign{
& \int d\theta d\varphi d\psi \sin\theta \,
D^{J_1}_{m_1 \, \mu_1} (\varphi, \theta, \psi)
D^{J_2}_{m_2 \, \mu_2} (\varphi, \theta, \psi)
D^{J_3}_{m_3 \, \mu_3} (\varphi, \theta, \psi) \cr
& \; = 2\pi \, \delta_{\mu_1 + \mu_2 + \mu_3, 0}
\int d\theta d\varphi \sin\theta \,
D^{J_1}_{m_1 \, \mu_1} (\varphi, \theta, 0)
D^{J_2}_{m_2 \, \mu_2} (\varphi, \theta, 0)
D^{J_3}_{m_3 \, \mu_3} (\varphi, \theta, 0) \cr
& \; = 8 \pi^2 \, \delta_{m_1 + m_2 + m_3, 0} \,
\delta_{\mu_1 + \mu_2 + \mu_3, 0}
\left( \matrix{
\hfill J_1 & \hfill J_2 & \hfill J_3 \hfill \cr
\hfill m_1 & \hfill m_2 & \hfill m_3 \hfill \cr
} \right)
\left( \matrix{
\hfill J_1 & \hfill J_2 & \hfill J_3 \hfill \cr
\hfill \mu_1 & \hfill \mu_2 & \hfill \mu_3 \hfill \cr
} \right),
}\eqn\intd
$$
where
$\left( \matrix{\hfill J_1 & \hfill J_2 & \hfill J_3 \hfill \cr
\hfill m_1 & \hfill m_2 & \hfill m_3 \hfill \cr} \right)$
is the $3j$-symbol.
Using eqs.\ \dd\ and \intd\ the effective action can be written as
$$
S_3^{(2)} = g_o\int d\phi \e^{2\phi} \int\nolimits_{S^2}
d\theta d\varphi \sin\theta \; {\rm Tr} \left( \Phi_0 \left[
\nabla_- \Phi_{1 \over 2},
\nabla_+ \Phi_{-{1 \over 2}} \right] \right).
\eqn\halfintegeraction
$$
The sum of eqs.\ \integeraction\ and \halfintegeraction\ gives the
complete cubic terms for the discrete states \cubic.
\par
We can also construct the effective action for the closed string.
Similarly to the open string case \threepoint, we can obtain the
three-point correlation function of discrete state
$$
\Psi^{(s)}_{J, m, m'}(z, \bar z)
= \Psi^{(s)}_{J, m}(z) \Psi^{(s)}_{J, m'}(\bar z).
\eqn\closedvertex
$$
The result can be summarized by the effective action.
Introducing a closed string variable $g^{(s)}_{J,m,m'}$ $(s=\pm )$
for each discrete state,
the cubic terms of the effective action determined by the
OPEs \discreteope\ are \NPrefmark{\klpo}
$$
\eqalign{
S_{3,c} = - & \, {g_c \over 2} \sum_{J_1, m_1, m'_1, J_2, m_2, m'_2}
(J_2 m_1-J_1 m_2)(J_2 m'_1-J_1 m'_2) \cr
& \times g^{(-)}_{J_1+J_2-1, -m_1-m_2, -m'_1-m'_2}
g^{(+)}_{J_1, m_1, m'_1} g^{(+)}_{J_2, m_2, m'_2}
 \int d\phi,
}\eqn\closedcubic
$$
where $g_c$ is the closed string coupling constant.
Following ref.\ \NPrefmark{\klpo} we define
a scalar field on $\R \times {\rm S}^3$
$$
\Phi_c (\phi, \theta, \varphi, \psi)
= \sum_{s, J, m, m'} g_{J, m, m'}^{(s)} n^s(J, m, m')
D^J_{m m'} (\varphi, \theta, \psi) \e^{(s J-1)\phi}.
\eqn\closedfield
$$
Using eq.\ \intd\ one obtains the closed string action
as \NPrefmark{\klpo}
$$
S_{3,c} = {g_c \over 3!}\int d\phi \e^{2\phi}
{1 \over 8\pi^2}\int\nolimits_{S^3} d\theta d\varphi
d\psi \sin\theta \; \Phi_c^3.
\eqn\closedaction
$$
In obtaining the above result, we find
the normalization factor in eq.\ \closedfield
$$
\eqalign{
n^{+}(J, m, m') &
= {\sqrt{(J+m)!(J-m)!(J+m')!(J-m')!} \over (2J-1)!}, \cr
n^{-}(J, m, m') & = (-1)^{m-m'+1}
{J(J+1)(2J+1)(2J-1)! \over \sqrt{(J+m)!(J-m)!(J+m')!(J-m')!}}.
}\eqn\closednorm
$$
%
%
\chapter{Effective Action involving Tachyons}
Since the noncompact case can be obtained by taking the infinite
radius limit, we can consider, without loss of generality,
a string compactified on a circle with a radius $R$
$$
X \sim X+2\pi R = X + 2\pi a\sqrt{\alpha'},
\eqn\radius
$$
where $a$ is the reduced radius in unit of the self dual radius
$\sqrt{\alpha'}$.
Both discrete momenta and winding numbers should be taken into
account in the case of the closed string.
We normalize discrete momenta $p$ and winding numbers $w$
$$
p={n \over 2a}, \quad w={ma\over 2}, \qquad  n, m \in \Z
\eqn\dwmomenta
$$
to take integer or half odd integer values at the self dual
radius $a=1$. It is also convenient to define the left and right
moving momenta as linear combinations (our definition is the usual
one divided by $\sqrt2$)
$$
p_L={p+w \over 2}, \qquad
p_R={p-w \over 2}.
\eqn\lrmomenta
$$
At the self-dual radius $a=1$, the discrete momenta and the
winding numbers are restricted to those corresponding to
discrete states.
In order to have couplings with the discrete states, we shall take
the radius of the compact boson to be a fractional multiple of
the self dual radius
$$
a={N \over M}, \qquad N, M \in \Z.
\eqn\fracrad
$$
Noncompact $X$ can be obtained simply by taking the limit of infinite
radius $R =\sqrt{\alpha'}N/M \rightarrow \infty$.
Hence we do not lose any generality.
\par
Here we shall discuss the open string case by taking only the right or
left moving part of the closed string.
As in the previous section we obtain the three-point functions
involving tachyons from the OPEs \tachyonope
$$
\eqalign{
\VEV{\Psi'^{(+)}_{p_1}(z_1) \, \Psi'^{(+)}_{p_2}(z_2) \,
\tilde\Psi'^{(+)}_{J_3, m_3}(z_3)}
& = {c^{(+)}(p_1; p_2; J_3, m_3) \over
(z_1-z_2)(z_2-z_3)(z_3-z_1)}, \cr
\VEV{\Psi'^{(-)}_{p_1}(z_1) \, \Psi'^{(-)}_{p_2}(z_2) \,
\tilde\Psi'^{(+)}_{J_3, m_3}(z_3)}
& = {c^{(-)}(p_1; p_2; J_3, m_3) \over
(z_1-z_2)(z_2-z_3)(z_3-z_1)},
}\eqn\tthreepoint
$$
where the coefficients, which are equal to the integrated
correlation functions are
$$
\eqalign{
c^{(+)}(p_1; p_2; J_3, m_3)
& = \delta_{1-p_1-p_2, J_3} \delta_{m_3, J_3-1} (-1)^{2J_2}
{\Gamma(1-2p_2) \over 2 \, \Gamma(2p_1)}, \cr
c^{(-)}(p_1; p_2; J_3, m_3)
& = \delta_{1+p_1+p_2, J_3} \delta_{m_3, 1-J_3} \cr
& \quad \times (-1)^{J_1(2J_1-1)+J_2(2J_2-1)+2J_1+1}
{\Gamma(1+2p_2) \over 2 \, \Gamma(-2p_1)}.
}\eqn\tachthreecoeff
$$
\par
{}From these correlation functions
we find that there are two terms in the action for a three point
coupling involving tachyons in the open string
$$
S_{T, \, o} = S_{T, \, o}^{(+)} +S_{T, \, o}^{(-)}.
\eqn\topsum
$$
The action for tachyons with positive chirality is given in terms of
variables $g_p^{(+)A}$, and
$$
\eqalign{
S_{T, \, o}^{(+)} & = {g_o \over 2} \sum_{A, B, C} f^{ABC}
\sum_{n_1 \in \Z} \sum_{J_3=1/2}^{\infty} (-1)^{2J_2+1}
{\Gamma(1-2p_2) \over 2 \, \Gamma(2p_1)}
g^{(+)A}_{p_1} g^{(+)B}_{p_2} g^{(+)C}_{J_3, J_3-1} \int d\phi, \cr
& \qquad p_1 = {n_1 \over 2a}, \qquad p_2 = -p_1-J_3+1.
}\eqn\posaction
$$
One should note that the tachyon coupling $g_o$ is the same as
the discrete state coupling in eq.\ \cubic.
The action for tachyons with negative chirality is given in terms of
variables $g^{(-)A}_p$, and
$$
\eqalign{
S_{T, \, o}^{(-)} & = {g_o \over 2} \sum_{A, B, C} f^{ABC}
\sum_{n_1 \in \Z} \sum_{J_3=1/2}^{\infty}
(-1)^{J_1(2J_1-1)+J_2(2J_2-1)+2J_1} \cr
& \quad \times {\Gamma(1+2p_2) \over 2 \, \Gamma(-2p_1)}
g^{(-)A}_{p_1} g^{(-)B}_{p_2} g^{(+)C}_{J_3, 1-J_3} \int d\phi, \cr
& \qquad p_1 = {n_1 \over 2a}, \qquad p_2 = -p_1+J_3-1.
}\eqn\negaction
$$
\par
Since the natural mode function for the tachyon is given by
the momentum eigenstate, we introduce a tachyon field for positive
and negative chirality
$$
\eqalign{
T^{A}(\phi, X) & = T^{(+)A}(\phi, X)+T^{(-)A}(\phi, X), \cr
T^{(+)A}(\phi, X)
& = \sum_{n \in \Z} {1 \over \sqrt{4\pi a}} \, \Gamma(1-2p)
\e^{ipX} \e^{(p-1)\phi} g^{(+)A}_{p={n \over 2a}}, \cr
T^{(-)A} (\phi, X)
& = \sum_{n \in \Z} {1 \over \sqrt{4\pi a}} \, (-1)^{J(2J-1)} \,
\Gamma(1+2p) \e^{ipX} \e^{(-p-1)\phi} g^{(-)A}_{p={n \over 2a}}.
}\eqn\topfield
$$
To write down the action we need another field constructed from the same
coefficients $g^{(\pm)A}_p$
$$
\eqalign{
\tilde T^{A}(\phi, X)
& = \tilde T^{(+)A}(\phi, X)+\tilde T^{(-)A}(\phi, X), \cr
\tilde T^{(+)A}(\phi, X)
& = \sum_{n \in \Z} {1 \over \sqrt{4\pi a}} \, (-1)^{2J-1} \,
\Gamma(1-2p) \e^{ipX} \e^{(p-1)\phi} g^{(+)A}_{p={n \over 2a}}, \cr
\tilde T^{(-)A} (\phi, X)
& = \sum_{n \in \Z} {1 \over \sqrt{4\pi a}} \, (-1)^{J(2J-1)+2J-1} \,
\Gamma(1+2p) \e^{ipX} \e^{(-p-1)\phi} g^{(-)A}_{p={n \over 2a}}.
}\eqn\tildetopfield
$$
By rewriting the variables $g_p^{(\pm )A}$ in terms of the field
$T^{(\pm )A}$ and $\tilde T^{(\pm )A}$, we can obtain the
coordinate representation of the tachyon action \topsum
$$
\eqalign{
S^{(+)}_{T, \, o} & = -{1 \over 8\pi} i g_o \sum_{A, B, C} f^{ABC}
\int d\phi \e^{2\phi} \int_{-2\pi a}^{2\pi a} dX \cr
& \quad \times \left[
T^{(+)A}(\phi, X+2\pi)-T^{(+)A}(\phi, X-2\pi) \right]
\tilde T^{(+)B}(\phi, X) \Psi^C (\phi, X), \cr
S^{(-)}_{T, \, o} & = -{1 \over 8\pi} i g_o \sum_{A, B, C} f^{ABC}
\int d\phi \e^{2\phi} \int_{-2\pi a}^{2\pi a} dX \cr
& \quad \times \left[
\tilde T^{(-)A}(\phi, X+2\pi)-\tilde T^{(-)A}(\phi, X-2\pi) \right]
T^{(-)B}(\phi, X) \Psi^C (\phi, X),
}\eqn\topaction
$$
where we have introduced the $X$ representation of the discrete states
$$
\eqalign{
\Psi^{A} (\phi, X) &=\sum_{s=\pm }\Psi^{(s)A} (\phi, X), \cr
\Psi^{(s)A} (\phi, X) &=\sum_{J=1/2}^{\infty}
\sum_{m=-J}^{J} \e^{imX} \e^{(sJ-1)\phi} g^{(s)A}_{J,m}.
}\eqn\xdfield
$$
Actually only the discrete states with $s=+$ and $m=\pm (J-1)$
contribute to the action.
\par
Now we shall discuss the closed string case by combining the left
and right moving part.
Since the Liouville zero mode is effectively noncompact for the bulk
amplitudes, the Liouville energy has to be common to left and right
movers.
If the left moving and right moving momenta are equal, the tachyon
corresponds to the center of mass motion and has discrete momenta $p$
but without winding numbers.
We shall denote the field for such a mode as $g^{(s,s)}_p$
with $s$ being the common chirality of left and right movers.
If the left moving and right moving momenta have opposite sign,
the tachyon has winding numbers $w$ but without discrete momenta.
We shall denote the field for such a mode as $g^{(s,-s)}_{w}$
with $s$ being the chirality of left movers.
\par
{}From the OPEs \tachyonope, we find that there are two terms
in the action for three point couplings involving tachyons
$$
S_{T, \, c} = S_{T, \, c}^{cm} +S_{T, \, c}^{wd}.
\eqn\tachsum
$$
The action for tachyons with discrete momenta (corresponding to
the center of mass motion) is given by
$$
\eqalign{
S_{T, \, c}^{cm} & = -{g_c \over 2} \sum_{n_1 \in \Z}
\sum_{J_3=1/2}^\infty \sum_{s=\pm}
\biggl[{\Gamma(1-2sp_2) \over 2 \, \Gamma(2sp_1)}\biggr]^2
g^{(s,s)}_{p_1} g^{(s,s)}_{p_2} g^{(+)}_{J_3, s(J_3-1), s(J_3-1)}
\int d\phi, \cr
& \qquad p_1 = {n_1 \over 2a}, \quad p_2 = -p_1-s(J_3-1).
}\eqn\mcmaction
$$
One should note that the tachyon coupling $g_c$ is the same coupling as
the discrete state coupling in eq.\ \closedcubic.
The action for tachyons with winding numbers is given by
$$
\eqalign{
S_{T, \, c}^{wd} & = {g_c \over 2} \sum_{n_1 \in \Z}
\sum_{J_2=1/2}^\infty \sum_{s=\pm }
(-1)^{J_1(2J_1+1)+J_2(2J_2+1)+2J_3} \cr
& \quad \times
\biggl[{\Gamma(1-2sp_2) \over 2 \, \Gamma(2sp_1)}\biggr]^2
g^{(s,-s)}_{p_1} g^{(s,-s)}_{p_2} g^{(+)}_{J_3, s(J_3-1), -s(J_3-1)}
\int d\phi, \cr
& \qquad p_1 = {n_1 a \over 2}, \quad p_2 = -p_1-s(J_3-1).
}\eqn\mwdaction
$$
\par
Exactly analogous to the open string case, we introduce the closed
string tachyon field for center of mass motion
$$
\eqalign{
T^{cm}(\phi, X) = & \, \sum_{s=\pm }T^{cm(s)}(\phi, X), \cr
T^{cm(s)}(\phi, X)
= & \, \sum_{n \in \Z}{1 \over \sqrt{4\pi a}}
{\Gamma (1-2sp) \over \Gamma (2sp)} \e^{ipX} \e^{(sp-1)\phi}
g^{(s,s)}_{p={n \over 2a}},
}\eqn\tcmfield
$$
where $X$ is the center of mass coordinate (multiplied by
$2/\sqrt{\alpha'}$).
Similarly we introduce the closed string tachyon field for the winding
modes
$$
\eqalign{
T^{wd}(\phi, X_r) = & \, \sum_{s=\pm }T^{wd(s)}(\phi, X_r), \cr
T^{wd(s)}(\phi, X_r)
= & \, \sum_{n \in \Z} \sqrt{{a \over 4\pi}} \,
(-1)^{J(2J+1)} \, {\Gamma (1-2sp) \over \Gamma (2sp)}
\e^{ip_rX_r} \e^{(sp_r-1)\phi} g^{(s,-s)}_{p_r={na \over 2}}.
}\eqn\twdfield
$$
We also need a field in the $X$ representation for the discrete
states of closed string
$$
\eqalign{
\Psi_c^{cm}(\phi, X)&=\sum_{s=\pm }\Psi_c^{cm(s)}(\phi, X), \quad
\Psi_c^{wd}(\phi, X_r)=\sum_{s=\pm }\Psi_c^{wd(s)}(\phi, X_r), \cr
\Psi_c^{cm(s)}(\phi, X)&=\sum_{J=1/2}^{\infty} \sum_{m=-J}^{J}
(-1)^{2J} \e^{imX} \e^{(sJ-1)\phi} g^{(s)}_{J,m,m}, \cr
\Psi_c^{wd(s)}(\phi, X_r)&=\sum_{J=1/2}^{\infty} \sum_{m=-J}^{J}
(-1)^{2J} \e^{imX_r} \e^{(sJ-1)\phi} g^{(s)}_{J,m,-m}.
}\eqn\xdiscretefield
$$
Actually only the discrete states with $s=+$ and $m=\pm (J-1)$
contribute to the action.
\par
{}From eq.\ \tachsum\ we find the coordinate representation of the
cubic terms in the effective action involving tachyons
$$
\eqalign{
S_{T, \, c}^{cm} & = {g_c \over 8}
\int d\phi \e^{2\phi} \int_{-2\pi a}^{2\pi a} dX
\, T^{cm}(\phi, X) \, T^{cm}(\phi, X) \, \Psi_c^{cm}(\phi, X), \cr
S_{T, \, c}^{wd} & = {g_c \over 8}
\int d\phi \e^{2\phi} \int_{-2\pi /a}^{2\pi /a} dX_r
\, T^{wd}(\phi, X_r) \, T^{wd}(\phi, X_r) \, \Psi_c^{wd}(\phi, X_r).
}\eqn\closedtachyonaction
$$
\par
We have succeeded to write the effective action involving tachyons
in a space with two flat variables $\phi$ and $X$ or $X_r$.
On the other hand, the effective action for discrete states
is written down in a space with $\R \times {\rm S}^2$
($\R \times {\rm S}^3$) for the open (closed) string.
It is possible to consider any convenient space to construct
the coordinate representation of an action.
However, it is certainly desirable to use the same space and the same
field for the same dynamical degree of freedom to write down various
pieces of an effective action.
Since the momentum conservation in the $X$ representation corresponds
precisely to the magnetic quantum number conservation in the angular
representation,  we have tried to use the angular coordinates
$\theta, \varphi$ and $\psi$ on the sphere to describe the tachyon
field. However, we were not quite successful, apparently because
the natural wave function of the tachyon is the momentum
eigenstate in the $X$ representation.
\par
\vskip 5mm
\ack
One of the authors (NS) thanks M. Sakamoto for a discussion on
cocycle factors, Y. Kitazawa for a useful discussion of the Liouville
theory. We would like to thank Patrick Crehan for a careful reading
of the manuscript. This work is supported in part by Grant-in-Aid
for Scientific Research from the Ministry of Education,
Science and Culture (No.01541237).
\vskip 5mm
\refout
\end
It has been observed that the cocycle operator multiplied by the zero
mode part of the vertex operator is very useful in showing the
cocycle property \NPrefmark{\gool}
$$
\hat c_\alpha \equiv \exp{i{\bf p}{\bf X_0}} c_\alpha
= \sum_{\beta \in \Lambda} \varepsilon (\alpha, \beta)
\ket{\alpha+\beta} \bra{\beta},
\eqn\hcalpha
$$
\par

Since the discrete states with $m=\pm J$ are actually tachyons, the
three tachyon coupling is actually contained in the action.

We find that the above OPEs exhaust all possible three point coupling
involving tachyons.
,
\cr
& \qquad\qquad\qquad\qquad\qquad\qquad (J_1 < J_3+1)